\documentclass[preprint,showpacs,preprintnumbers,amsmath,amssymb, showkeys]{revtex4}

\usepackage{array}
\usepackage{booktabs}
\usepackage{tabu}
\usepackage{dcolumn}
\usepackage{amsmath}
\usepackage{amsfonts}
\usepackage{amssymb}
\usepackage{graphicx}
\usepackage{subfigure}
\usepackage{graphicx}
\usepackage{dcolumn}
\usepackage{bm}

\def\be{\begin{equation}}
\def\ee{\end{equation}}
\def\bea{\begin{eqnarray}}
\def\eea{\end{eqnarray}}
\def\f{\frac}
\def\n{\nonumber}
\def\l{\label}
\def\p{\phi}
\def\o{\over}
\def\R{\rho}
\def\pa{\partial}
\def\om{\omega}
\def\na{\nabla}
\def\P{\Phi}

\begin{document}

\title{Anisotropic inflation in Brans-Dicke gravity}

\author{M. Tirandari}
\email{mehraneh.tirandari@gmail.com}
\author{Kh. Saaidi}
 \email{ksaaidi@uok.ac.ir}
\affiliation{Department of Physics, University of Kurdistan, \\Pasdaran St., P.O. Box 66177-15175, Sanandaj, Iran}

\def\be{\begin{equation}}
\def\ee{\end{equation}}
\def\bea{\begin{eqnarray}}
\def\eea{\end{eqnarray}}
\def\f{\frac}
\def\n{\nonumber}
\def\l{\label}
\def\p{\phi}
\def\o{\over}
\def\R{\rho}
\def\pa{\partial}
\def\om{\omega}
\def\na{\nabla}
\def\P{\Phi}

\begin{abstract}
We study anisotropic inflation in the Brans-Dicke gravity in the presence of an abelian gauge field where the gauge field is non-minimally coupled to the inflaton. We show that the degree of anisotropy, under slow-roll approximations, is proportional to slow-roll parameter of the theory. As a demonstration, we consider the displaced quadratic potential for the inflation. We do the numerical calculation of the model to investigate the behavior of anisotropy by changing the parameter in the Brans-Dicke model. We find out that, the solution is an attractor in the phase space, and anisotropy grows with the number of e-folds. Anisotropy depends on the Brans-Dicke parameter, $ \omega $, initial values of the scalar field and constant parameter of the coupling function of the scalar field and the abelian gauge field, $ c $. If we consider upper bound on the number of e-folds from CMB i.e. $ 60 $ e-folds, by increasing $ \omega $ and $ c $, anisotropy do not have time to exit the horizon and it is suppressed. 
\end{abstract}
\maketitle

\section{introduction}
\label{sec:intro}
Solving the problem of hot Big-Bang theory, and providing a natural mechanism for generating primordial density perturbations, make the elegant scenario of inflation as one of the best candidate of very early universe. The scenario has opened a new and unique window on physics at inaccessible energy scales\cite{Guth}-\cite{Linde}.
During inflation, quantum vacuum fluctuations are redshifted far outside the Hubble radius, imprinting spectrum of classical density perturbations \cite{Aghamohammadi}, \cite{Golanbari}.

Inflation can be well-defined by a quasi de sitter expansion in which the de Sitter expansion is not accurate. Because of this, statistics of cosmic microwave background (CMB) fluctuation is nearly Gaussian, also the power spectrum of fluctuations is almost scale invariant  and almost statistically isotropic. Therefore, we have to look at fine structures of the CMB fluctuation, such as non-Gaussianity which is a result of violation of translational symmetry and doesn't appear in the simplest single field inflaton with a canonical action \cite{Komatsu}-\cite{Bartolo}, spectral tilt as a result of violation of the temporal part of the de Sitter symmetry which is characterized by a slow-roll parameter, and statistical anisotropy which means spatial part of the de Sitter symmetry is not exact \cite{soda1}. Although deviation from Gaussianity, scale invariance and statistical isotropy are quite small, they are observationally corroborated by Planck and WMAP \cite{Akrami}-\cite{Planck 2015}.
 
Some anomalies in the CMB are of great importance \cite{Bennett}, \cite{Schwarz}. They have been noted since the early WMAP releases \cite{Eriksen}. They have been studied extensively by independent groups, and remains unresolved to the present day \cite{Eriksen2007}, \cite{Hanson}. Among them, there is hemispherical power asymmetry (dipolar or quadrupole power asymmetry).  Despite the fact that the statistical significance of hemispherical power asymmetry is hard to quantify (especially for the case quadrupole), researchers has been prompting to give variety of models, which have been put forward to explain this phenomenon \cite{Berera}-\cite{Jazayeri}.

The first attempt to put constraints on a preferred direction during inflation, and  to make robust predictions for its observable consequence and its magnitude, was done by Ackerman \textit{et al.}  \cite{Ackerman}. They considered the most general form of the power spectrum (in which rotational invariance has violated) as follows,
 \begin{equation}
\label{eq:emc}
P(\textbf{k}) = P(k)(1 + g_{*}({\textbf{k}}.\textbf{n})^{2}).
\end{equation}
Here $P(k)$ is the power spectrum for the primordial density perturbations $ \delta(k) $ and depends only on the magnitude of the vector \textbf{k}. Also, $ \textbf{n} $ is the privileged direction by which rotational invariance is broken. Moreover, $ g_{*} $ characterizes the deviation from the isotropy. The obtained bound using WMAP data is $ g_{*}=0.29\pm0.031$ \cite{Groeneboom}. A later analysis based on Planck data gave the best constraint i.e. $ g_{*}=0.002\pm0.016 (68 \% CL) $ so far \cite{Kim and Komatsu}. The Planck team got very similar constraints \cite{Planck 2015}.

An inflationary theory can accommodate models that produce hemispherical power asymmetry. Anisotropic inflation is categorized as multifield inflation models \cite{Mukhanov}. It contains a vector field leading to a breaking of statistical isotropy. So the power spectrum for the primordial density perturbations becomes statistically anisotropic. In this model the inflaton field $ \phi $ is coupled to the gauge kinetic term in the form $ (-f(\phi)F_{\mu\nu}F^{\mu\nu})/4 $ in the Bianchi type I (BI) space. There exists an attractor behavior for the solution and the vector field survives during inflation. Also, it is free from the ghost \cite{soda}. This model has been studied extensively in the literature \cite{emami}-\cite{lahiri}. Breaking the rotational invariance during inflation, in which we call it anisotropy, is characterized by the background shear divided by the Hubble rate. It is shown in \cite{komatsu} that for the bound from Planck, anisotropy is less than $ \mathcal{O}(10^{-9}) $.  However, they have shown that  for reasonable values of parameters of the model, which they are consistent with observation, it is hard to reach the attractor solution during inflation. Nevertheless, one can assume that the duration of inflation exceeds $ 60 $ efolds, the model settles down to its attractor solution before perturbations exit the horizon.

The prototype of an alternative to Einstein's general relativity was done by Brans and Dicke \cite{Brans}. The primary motivation for their theory comes from Mach's principle, that the phenomenon of inertia ought to arise from accelerations with respect to the general mass distribution of the universe \cite{weinberg}. Brans-Dicke (BD) theory is an important branch of the extended theories of gravity in the scalar-tensor theories. In BD theory, however, the gravitational coupling is variable. It is determined by all matter in the universe, accordingly, a scalar field is considered to couple to the Ricci curvature nonminimally. In spite of declining of interest in BD gravity in the $ 1970s $, a surge interest has raised owing to the new importance of scalar fields in unified theories, in particular string theory. Another reason for this interest is discovering plausible mechanisms that allow the parameter $ \omega $ (a variable in the BD gravity) to get values of order unity in the early universe and diverge later \cite{Faraoni}. Finally, the using of scalar tensor gravity theories in inflationary scenarios of the universe, has renewed interest in BD gravity \cite{Fujii}-\cite{Karami}.

This paper is organized as follows. In Section \ref{S:2}, the general action of anisotropic inflation in scalar tensor gravity is provided. Then, the most general equations of motion is obtained. In Section \ref{S:3},  anisotropic Brans-Dicke inflation is considered, and anisotropy is obtained in the terms of the slow roll parameters. Also, a relation for the phase transition is obtained. In Section \ref{S:4}, numerical calculation is performed for a specific potential. It shows that, there is an attractor solution, phase transition occurs and anisotropy grows in this model. Conclusion remarks are given in Section \ref{conclusion}.

\section{Anisotropic inflation in Scalar-Tensor gravity}
\label{S:2}

 In order to generate anisotropic effect during inflation, we add an abelian massless gauge kinetic term. It is coupled to the inflaton field through a gauge coupling function $ f^{2}(\phi) $, to the general action of scalar tensor models. So the action is given by

 \begin{equation}
\label{SscalarTensor}
S=\int d^{4}x\sqrt{-g} \left[\frac{1}{2} \mathcal{F}(R,\phi)-\frac{1}{2}\omega(\phi)g^{\mu\nu}\partial_{\mu}\phi\partial_{\nu}\phi-U(\phi)-\frac{1}{4}f^{2}(\phi)F_{\mu\nu}F^{\mu\nu} \right], 
\end{equation}
where $ \mathcal{F}(R,\phi) $ is a general function of the Ricci scalar $ R $ and the scalar field $ \phi $. Moreover, $ \omega(\phi) $ and $ U(\phi) $ are functions of $ \phi $, and $ F_{\mu\nu}\equiv\partial_{\mu} A_{\nu}-\partial_{\nu} A_{\mu} $ is the field strength tensor of the gauge field $ A_{\mu} $. Also, $ f(\phi) $ is the coupling function between $ \phi $ and $ A_{\mu} $. It will be specified later. We have also set the Planck scale $ M_{p}=1 $ for convenience. We focus on the BI metric, given by
 \begin{equation}
\label{BImetric}
ds^{2}=-dt^{2}+e^{2\alpha(t)-4\sigma(t)}dx^{2}+e^{2\alpha(t)+2\sigma(t)}(dy^{2}+dz^{2}),
\end{equation}
where $ a\equiv e^{\alpha(t)} $ is the isotropic scale factor and $ \sigma(t) $ is spatial shear which represents deviation from the isotropy. We choose the gauge $ A_{0}=0 $. Moreover, x-axis provides direction of the gauge field, without loss of  generality. Hence, we have the homogeneous gauge field of the form $ A_{\mu}=(0,A(t), 0, 0) $. We take the scalar field to be homogeneous of the form $ \phi(t) $ too. Tacking variation of action \eqref{SscalarTensor} with respect to the dynamical variable leads to the following equations

\begin{equation*}
\frac{1}{2}\mathcal{F}(R,\phi)g_{\mu\nu}-\left[R_{\mu\nu}+g_{\mu\nu}\bigtriangledown_{\lambda}\bigtriangledown^{\lambda}-\bigtriangledown_{\mu}\bigtriangledown_{\nu}\right]  \frac{\partial\mathcal{F}(R,\phi)}{\partial R}- 
 g_{\mu\nu}U(\phi)+
 \end{equation*}
 \begin{eqnarray}
 \omega(\phi)\left[ \bigtriangledown_{\mu}\phi\bigtriangledown_{\nu}\phi-\frac{1}{2}g_{\mu\nu}\bigtriangledown_{\lambda}\phi\bigtriangledown^{\lambda}\phi \right] +
 \frac{1}{2}f^{2}(\phi)\frac{\partial(F_{\lambda\rho}F^{\lambda\rho})}{\partial g_{\mu\nu}}-\frac{1}{4}g_{\mu\nu}f^{2}F_{\lambda\rho}F^{\lambda\rho}&=&0
\label{1}\\
 \bigtriangledown_{\lambda}\bigtriangledown^{\lambda}\phi+\frac{1}{2\omega(\phi)}\left(\omega^{\prime}(\phi)\bigtriangledown_{\lambda}\phi\bigtriangledown^{\lambda}\phi-2U^{\prime}(\phi)+\frac{\partial \mathcal{F}(R,\phi)}{\partial\phi}+f (\phi) f^{\prime}(\phi)F_{\lambda\rho}F^{\lambda\rho}\right) &=&0
\label{2}
\\
\bigtriangledown_{\mu}\left( f^{2}(\phi)F^{\mu\nu}\right)&=&0
\label{3}\end{eqnarray}
where $ \bigtriangledown_{\mu} $ represents a covariant derivative with respect to the metric $ g_{\mu\nu} $ and a prime denotes a derivative with respect to $ \phi $. Using the metric \eqref{BImetric}, equation of motion of the gauge field \eqref{3} can be easily solved as 
\begin{equation}
\dot{A}_{x}=\frac{p_{A}}{f^{2}(\phi)e^{\alpha+4\sigma}},
  \label{4}\end{equation}
where an overdot denotes the derivative with respect to the cosmic time $ t $ and $ p_{A} $ denotes constant of integration. From equations \eqref{1}, \eqref{2} and using equation \eqref{BImetric}, we obtain constraint, evolution and inflaton field equations in the BI space as
\begin{eqnarray}\label{5}
(3 \dot{\alpha}^{2}-3\dot{\sigma} ^{2})Q+\frac{1}{2}(\mathcal{F}-RQ)+3\dot{\alpha} \dot{Q}-\frac{1}{2}\omega \dot{\phi}^{2}-U-\frac{1}{2}f^{2}\dot{A}_{x}^{2}e^{-2\alpha+4\sigma}&=&0,
\\
\label{6}
(\ddot{\alpha}+ 3 \dot{\alpha}^{2})Q+\frac{1}{2}(\mathcal{F}-RQ)+\frac{5}{2}\dot{\alpha}\dot{Q}+ \ddot{Q}-U-\frac{1}{6}f^{2}\dot{A}_{x}^{2}e^{-2\alpha+4\sigma}&=&0,
\\
\label{66}
(\ddot{\sigma}+ 3 \dot{\alpha}\dot{\sigma})Q+\dot{\sigma} \dot{Q}-\frac{1}{3}f^{2}\dot{A}_{x}^{2}e^{-2\alpha+4\sigma}&=&0,
\\
\label{7}
\ddot{\phi}+ 3\dot{\alpha}\dot{\phi}+\frac{1}{2\omega}\left( \omega^{\prime}\dot{\phi}^2-\frac{\partial\mathcal{F}}{\partial\phi}+2U^{\prime}- 
 2f^{3}f^{\prime}\dot{A}_{x}^{2}e^{-2\alpha+4\sigma}\right)&=&0 ,
 \end{eqnarray}
where $ Q $ is defined as $ Q\equiv \partial\mathcal{F}/\partial R $ and the Hubble parameter can be expressed as $ H\equiv \dot{\alpha} $. Eqs. \eqref{4}-\eqref{7} are the most general equations for the anisotropic scalar tensor inflationary model. When the abelian gauge field goes to zero, i.e. $ f\rightarrow 0 $, and $ \sigma\rightarrow 0 $ (i.e. spatially flat Friedmann-Robertson-Walker (FRW) universe), the model is reduced to that of the \cite{Tsujikawa}, \cite{Karami} and \cite{Myrzakulov}. Moreover, in the limit $ \mathcal{F}(R,\phi)=R $ and $ \omega(\phi)=1 $ it is reduced to \cite{soda}.

The slow-roll parameters which has been introduced for this model are \cite{Tsujikawa}, \cite{Hwang}
  \begin{equation} 
\varepsilon_{1}\equiv-\frac{\dot{H}}{H^{2}},\  \ \varepsilon_{2}\equiv-\frac{\ddot{\phi}}{H\dot{\phi}},\  \  \varepsilon_{3}\equiv-\frac{\dot{Q}}{2HQ},\  \  \varepsilon_{4}\equiv-\frac{\dot{E}}{2HE},
 \label{slowroll}\end{equation}
where the parameter $ E $ is defined as 
  \begin{equation}
E\equiv Q \left[ \omega(\phi)+\frac{3\dot{Q}^{2}}{2\dot{\phi} Q}\right].
\end{equation}
\section{Anisotropic inflation in Brans-Dicke gravity}
\label{S:3}
So far we have discussed the anisotropic scalar tensor inflationary model. In this section, we consider anisotropic BD inflation as a special case of the anisotropic scalar tensor inflationary model. The action in the so-called Jordan frame is 
   \begin{equation}
S=\int d^{4}x\sqrt{-g} \left[\frac{1}{2} \phi R-\frac{1}{2}\frac{\omega_{BD}}{\phi}g^{\mu\nu}\partial_{\mu}\phi\partial_{\nu}\phi-U(\phi)-\frac{1}{4}f^{2}(\phi)F_{\mu\nu}F^{\mu\nu}\right], 
 \label{BransDicke}\end{equation}
where this action is obtained from the action \eqref{SscalarTensor} by considering
   \begin{equation}
\mathcal{F}(R,\phi)=\phi R, \  \             
 \omega(\phi)=\frac{\omega_{BD}}{\phi},
 \label{eq}\end{equation}
where $ \omega_{BD} $ is the BD parameter which is a constant, and hereafter we drop out its subscript and write it as $ \omega $. Using Eqs. \eqref{4}-\eqref{7} for the action \eqref{BransDicke}, constraint, evolution and inflaton field equations are obtained in the BI universe, as
\begin{eqnarray}
 (\dot{\alpha}+\frac{\dot{\phi}}{2\phi})^{2}-\dot{\sigma} ^{2}-\frac{1}{3\phi}\left( (\frac{2\omega+3}{4})\frac{\dot{\phi}^{2}}{\phi}+U+\frac{1}{2}f^{-2}p_{A}^{2}e^{-4\alpha-4\sigma}\right) &=&0,
\label{8}
\\
\left( \ddot{\alpha}+ 3 \dot{\alpha}^{2}\right) +\frac{1}{\phi}\left( \frac{5}{2}\dot{\alpha}\dot{\phi}+\ddot{\phi}-U-\frac{1}{6}f^{-2}p_{A}^{2}e^{-4\alpha-4\sigma} \right)&=&0,
 \label{16}
 \\
(\ddot{\sigma}+ 3 \dot{\alpha}\dot{\sigma})+\frac{1}{3\phi}\left(3\dot{\sigma} \dot{\phi}-f^{-2}p_{A}^{2}e^{-4\alpha-4\sigma}\right)&=&0,
  \label{9}
  \\
\ddot{\phi}+ 3\dot{\alpha}\dot{\phi}+\frac{\phi}{2\omega}\left(-\omega \left( \frac{\dot{\phi}}{\phi}\right) ^{2}-R+2U^{\prime}-2f^{-3}f^{\prime}p_{A}^{2}e^{-4\alpha-4\sigma}\right)&=&0.
\label{scalareq}\end{eqnarray}
Where the last equation can be written in the form
\begin{equation}
\ddot{\phi}+ 3\dot{\alpha}\dot{\phi}+\frac{2}{2\omega+3}\left(\phi U^{\prime}-2U+3f^{-2}p_{A}^{2}e^{-4\alpha-4\sigma}-\phi f^{-3}f^{\prime}p_{A}^{2}e^{-4\alpha-4\sigma}\right)  =0.
 \label{11}\end{equation}
Using equations \eqref{6}, \eqref{8} and \eqref{11}, the equation for acceleration of the universe is given by
\begin{equation}
 \begin{split}
\ddot{\alpha}+\dot{\alpha}^{2}=\left( \frac{9-2\omega}{6}\right) \left( \frac{\dot{\phi}}{2\phi}\right) ^{2}+\frac{5}{2}\frac{\dot{\alpha}\dot{\phi}}{\phi}-9\dot{\sigma}^{2}+\left( \frac{1}{3}-\frac{4}{2\omega+3}\right) \frac{U}{\phi}+&\\ \left( \frac{6}{2\omega+3}-\frac{1}{6}\right) 2\frac{\rho_{A}}{\phi}+\frac{2}{2\omega+3}U^{\prime}-\frac{2}{2\omega+3}f^{-1}f^{\prime}2\rho_{A}.
 \end{split}
 \label{111}\end{equation}
The necessary condition for inflation is $ \ddot{a}/a=(e^{\alpha})^{\ddot{•}}/(e^{\alpha})\geq0 $, and it will be satisfied if the $ U $ term overcomes shear $ \Sigma=\dot{\sigma} $, energy density of the gauge field $ \rho_{A} =f^{-2}p_{A}^{2}e^{-4\alpha-4\sigma}/2 $ and kinetic  energy of the inflaton $ \dot{\phi}^{2}/2 $.  From equation \eqref{9}, it is clear that anisotropy will grow only when the last term i.e. $ \rho_{A} $ is a dominant term. Therefore if we consider $ \sigma\ll\alpha $, anisotropy starts to grow at least for $ f=e^{-2\alpha} $. A generalized form for $ f(\phi) $ is 
\begin{equation}
f(\phi)=e^{-2c\alpha},
 \label{coupling1}\end{equation}
 where $ c>1 $ is a constant. Then, the energy density of the gauge filed, ignoring $ \sigma $, is $ \rho_{A} =p_{A}^{2}e^{4(c-1)\alpha}/2 $.
Considering the slow roll conditions $ \vert\dot{\phi}\vert\ll\vert H\phi\vert $ and $ \vert\ddot{\phi}\vert\ll\vert3H\dot{\phi}\vert $, and additionally $ \sigma\ll\alpha $, $ \dot{\sigma}\ll\dot{\alpha} $ hold, so \eqref{8} and \eqref{scalareq} reduce to
\begin{eqnarray}
3\phi\dot{\alpha}^{2}&\simeq & U(\phi),
  \label{12}\\
3\dot{\phi}\dot{\alpha}&\simeq &\frac{2}{2\omega+3}\left[ 2U(\phi)-\phi U^{\prime}(\phi)\right].
  \label{SR1}\end{eqnarray}
Substituting above equation into Eq. \eqref{slowroll}, one arrives at the following expression for slow-roll parameters 
\begin{eqnarray}
\varepsilon_{1}&=&\frac{(U-\phi U^{\prime})(2U-\phi U^{\prime})}{(2\omega+3)U^{2}},
 \label{SR2}
 \\
\varepsilon_{2}&=&\varepsilon_{1}+\frac{2\phi (U^{\prime}-\phi U^{\prime \prime})}{(2\omega+3)U},
 \label{SR3}
 \\
\varepsilon_{3}&=&\frac{2U-\phi U^{\prime}}{(2\omega+3)U}=\frac{U\varepsilon_{1}}{(U-\phi U^{\prime})},
 \label{SR4}\end{eqnarray}

and $ \varepsilon_{4}=0 $. The number of e-folds which measures the amount of inflation, is given by
\begin{equation}
N=\alpha\equiv\int^{t}_{t_{end}}Hdt=\frac{2\omega +3}{2}\int \frac{U}{\phi(2U-\phi U^{\prime})}d\phi .
  \label{SR5}
\end{equation}
From\eqref{coupling1} the coupling function $ f(\phi) $  is 
\begin{equation}
f(\phi)=e^{-c(2\omega +3)\int \frac{U}{\phi(2U-\phi U^{\prime})}d\phi}.
\label{coupling2}\end{equation}
We are interested in the configuration where inflation take place with small anisotropies such that $ \Sigma/H\ll 1 $ but not zero. From equation \eqref{9} in the slow-roll approximation we have
\begin{equation}
\frac{\Sigma}{H}\simeq \left( \frac{3(2\omega+3)p_{A}^{2}}{(2\omega+3)3U+2(2U-\phi U^{\prime})}\right) g(\alpha)^{-1},
  \label{anisotropy}\end{equation}
where $ g(\alpha) $ is
\begin{equation}
g(\alpha)=f^{2}e^{4\alpha+4\sigma}.
\label{galpha}\end{equation}
It can be obtained from \eqref{scalareq} in the slow-roll approximation as
\begin{equation}
g(\alpha)=g(\alpha_{0})e^{-4(c-1)(\alpha-\alpha_{0})}+\frac{\Omega(\phi)}{4(c-1)},
  \label{galpha2}\end{equation}
where
\begin{equation}
\Omega(\phi)=4cp_{A}^{2}\left(\frac{(3(2U-\phi U^{\prime})+c(2\omega+3))U}{(2U-\phi U^{\prime})^{2}} \right).
\label{omega}\end{equation}
In the limit of $ \alpha\rightarrow -\infty $ the first term of equation \eqref{galpha2} dominated and $ g(\alpha) $ diverges to infinity. Consequently anisotropy goes to zero, i.e. $ \Sigma/H\rightarrow 0 $. On the other hand, in the limit $ \alpha\rightarrow \infty $, the second term is dominated, and $ g(\alpha)\rightarrow\Omega/4(c-1) $. Therefore 
\begin{eqnarray}
\frac{\Sigma}{H}&\rightarrow & \left( \frac{3(2\omega+3)p_{A}^{2}}{(2\omega +3)3U+2(2U-\phi U^{\prime})}\right) \frac{4(c-1)}{\Omega}\\ 
&=&\frac{3(c-1)}{c}\frac{(2\omega+3)\varepsilon_{3}^{2}}{(2\varepsilon_{3}+3)(3\varepsilon_{3}-c)}.
  \label{anisotropy3}\end{eqnarray}
 It is clear that there is a "phase transition" when the anisotropy $ \Sigma/H $ has a transient from zero to the above equation. If $ \varepsilon_{3} $ goes to zero, then $ \Sigma/H\rightarrow 0 $. 
  \section{Numerical analysis of Anisotropic inflation in Brans-Dicke gravity}
\label{S:4}

In order to make the analysis more precise, we consider displaced quadratic inflationary potential as follows 
 \begin{equation}
 \begin{split}
U(\phi)=\dfrac{1}{2}m^{2}(\phi-\phi_{0})^{2},
  \label{potential}\end{split}
\end{equation}
where $ m=10^{-5} $ and $ \phi_{0} $ is a shift in the potential. This potential is a generalized version of the Starobinsky $ R^{2} $ inflation in the Einstein frame. Consistency of this potential with the Planck $ 2015 $ data in BD gravity has been investigated by \cite{Karami} using Jordan frame  and with Planck $ 2013 $ data by \cite{Tsujikawa2} in the Einstein frame. The slow-roll parameters for this potential will become
 \begin{eqnarray}
\varepsilon_{1}&=&\dfrac{1}{2}\frac{(\phi+\phi_{0})\phi_{0}}{2\omega+3},
  \label{SR1}
\\
\varepsilon_{2}&=&\frac{(\phi^{2}-\phi_{0}^{2})\phi_{0}-4\phi \phi_0}{2(2\omega+3)(\phi-\phi_{0})^{2}},
  \label{SR2}
\\
\varepsilon_{3}&=&-\frac{(\phi-\phi_{0})\phi_{0}}{2\omega+3}.
  \label{SR3}
  \end{eqnarray}
Therefore from equation \eqref{SR3}, as $ \phi\rightarrow\phi_0 $, the parameter $ \varepsilon_{3} $ approaches zero and from \eqref{anisotropy3} $ \Sigma/H\rightarrow0 $.
\subsection{Phase plane}
\label{S:4.1}

\begin{figure}[tbp]
\centering 
\includegraphics[width=.7\textwidth]{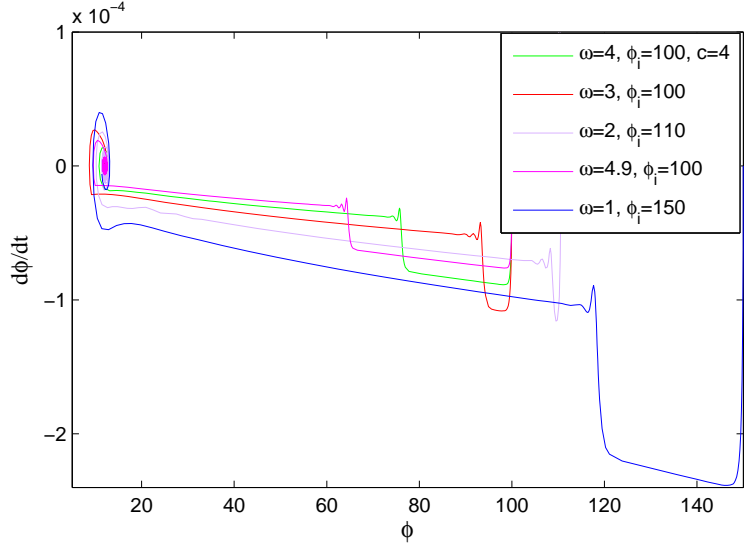}
\caption{\label{phase} Phase flow for $ \phi $ is depicted, $ c=4 $,  $ \omega =1$ to $4.9 $, initial conditions $ \alpha_{i}=\sigma_{i}=\dot{\sigma}_{i}=\dot{\phi}_{i}=0 $ and $ \phi_0=12 $ is considered. There is two slow-roll phases.}
\end{figure}

\begin{figure}[tbp] 
 \includegraphics[width=.7\linewidth]{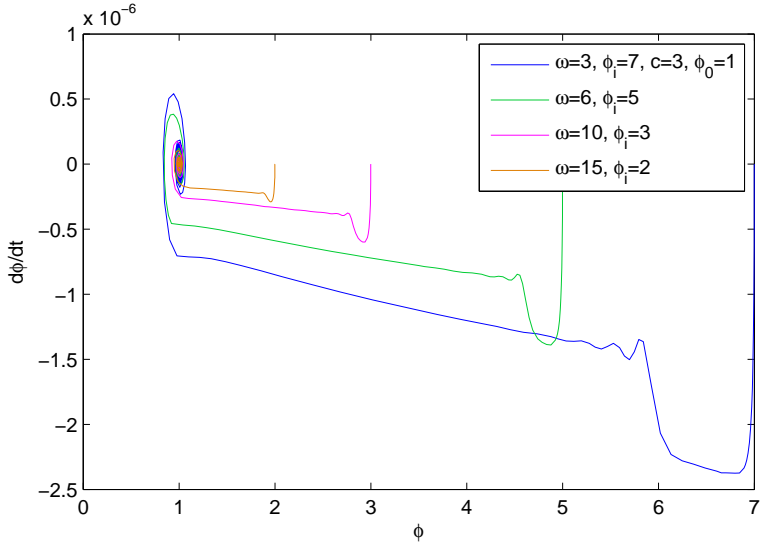} \centering
   \caption{ The same as Fig. \ref{phase}, but for $ \phi_0=1 $, constant value of $ c=3 $ and different values of $ \omega $. }
   \label{phasePhi01r3w3to15}
    \end{figure}

Solving eqs. \eqref{8}-\eqref{scalareq} numerically for the potential \eqref{potential} with different values of $ \omega $ and considering $ \phi_0=12 $, phase-plane in $ \dot{\phi}-\phi  $ is obtained (Fig. \ref{phase}). Obviously, there are two slow-roll inflationary phases, a conventional isotropic inflationary phase and an anisotropic inflationary phase. It is clear that, as we increase $ \omega $, the phase of anisotropic inflation starts later near the end of inflation. Moreover, by decreasing $ \omega $, anisotropic inflation starts sooner and it lasts more time until the end of inflation. The same plot for $ \phi_0=1 $ has been shown in Fig. \ref{phasePhi01r3w3to15}. Clearly, by increasing $ \omega $, we have to decrease $ \phi_i $ to have attractor solution. From equation \eqref{anisotropy3}, if $ \phi_i\rightarrow\phi_0 $ (i.e. $ \varepsilon_{3}\rightarrow0 $), then anisotropy, $ \Sigma/H $ goes to zero. This fact is consistent with Fig. \ref{phasePhi01r3w3to15}.

\subsection{Anisotropy}
\label{S:4.2}
\begin{figure}[tbp] 
 \includegraphics[width=.7\linewidth]{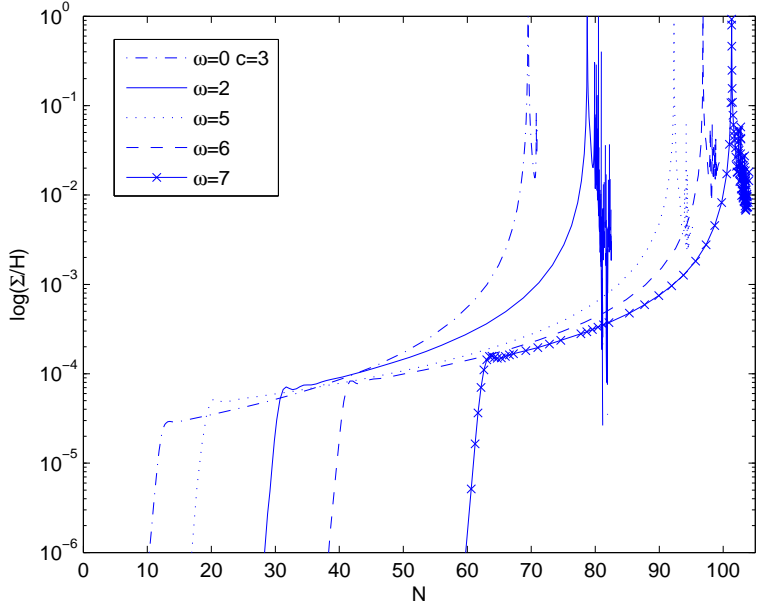} \centering
   \caption{Evolution of the anisotropy parameter $ \Sigma/ H $ with respect to the e-folding number $ N $, for different values of $ \omega  $  and $ c=3 $.  Other initial values are the same as in Fig. \ref{phase}. }
   \label{2anisotropyOmega0to7r3}
    \end{figure}

Evolution of the anisotropy parameter $ \Sigma/ H $ with respect to the e-folding number $ N $, for $ \omega =0$ to  $ 7 $, $ c=3 $ and $ \phi_0=12 $ is presented in Fig. \ref{2anisotropyOmega0to7r3}. Obviously, by increasing $ \omega $,  we have to increase $ \phi_{i} $ in order to have attractor solution. Thus, anisotropy is shifted out of the $ 60 $ e-folds number. Therefore, anisotropy dose not have time to exit the horizon and it is suppressed during inflation. 

It should be noticed that the value of BD parameter $ \omega $ is small and it is not compatible with local gravity tests in the solar system (i.e. $ \omega>40000 $). But this lower bound was made for pure Brans-Dicke theory, and these two models can't be compared to each other. Of course, during the slow roll inflation, kinetic energy of the inflaton is smaller than the potential energy. It seems that  $ \omega $ is small during this era and, the BD scalar field decays after inflation. Thus, the subsequent cosmology coincides with Einstein's general relativity and todays bound is invalid during inflation.
    \begin{figure}[tbp]
 \includegraphics[width=.7\linewidth]{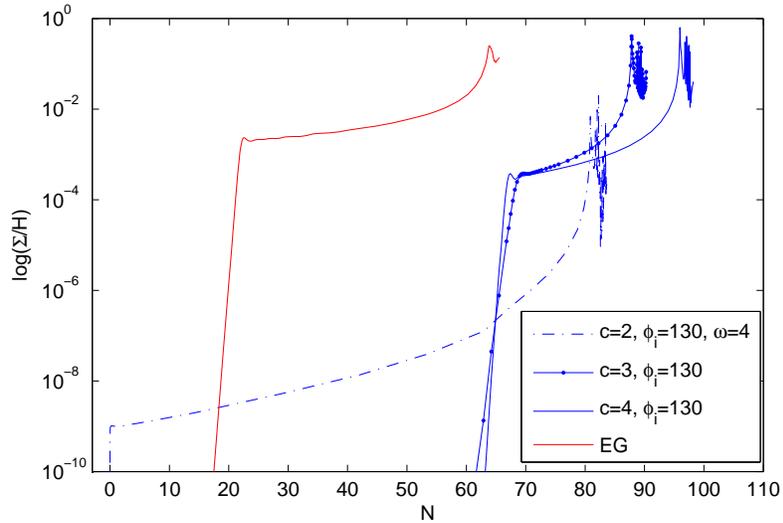} \centering
   \caption{The same as Fig. \ref{2anisotropyOmega0to7r3}, but for a constant value of $ \omega=4 $ and variable value of $ c $. The red line is Anisotropy in the Einstein gravity for the same potential.}
   \label{anisotropyOmega4r2to45phi100to130}
    \end{figure}
        \begin{figure}[tbp]
 \includegraphics[width=.7\linewidth]{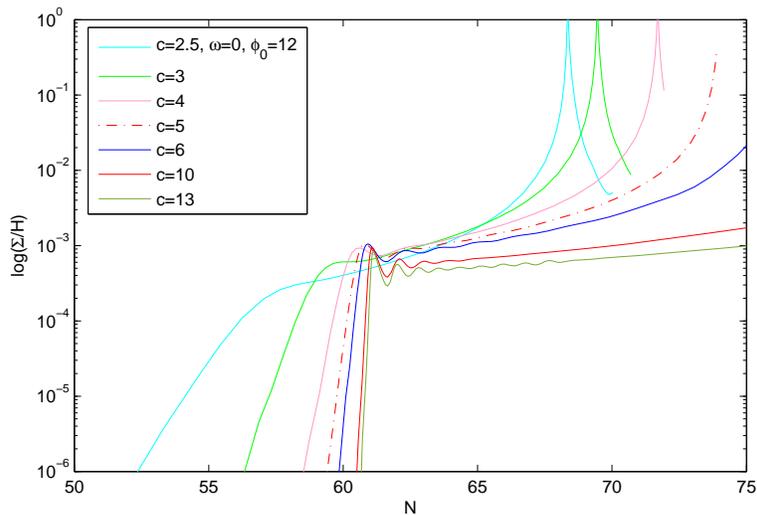} \centering
   \caption{The same as Fig. \ref{2anisotropyOmega0to7r3} but for a constant value of $ \omega=0 $ and different values of $ c $. }
   \label{3anisotropyOmega0r25to13}
    \end{figure}
      \begin{figure}[tbp] 
 \includegraphics[width=.7\linewidth]{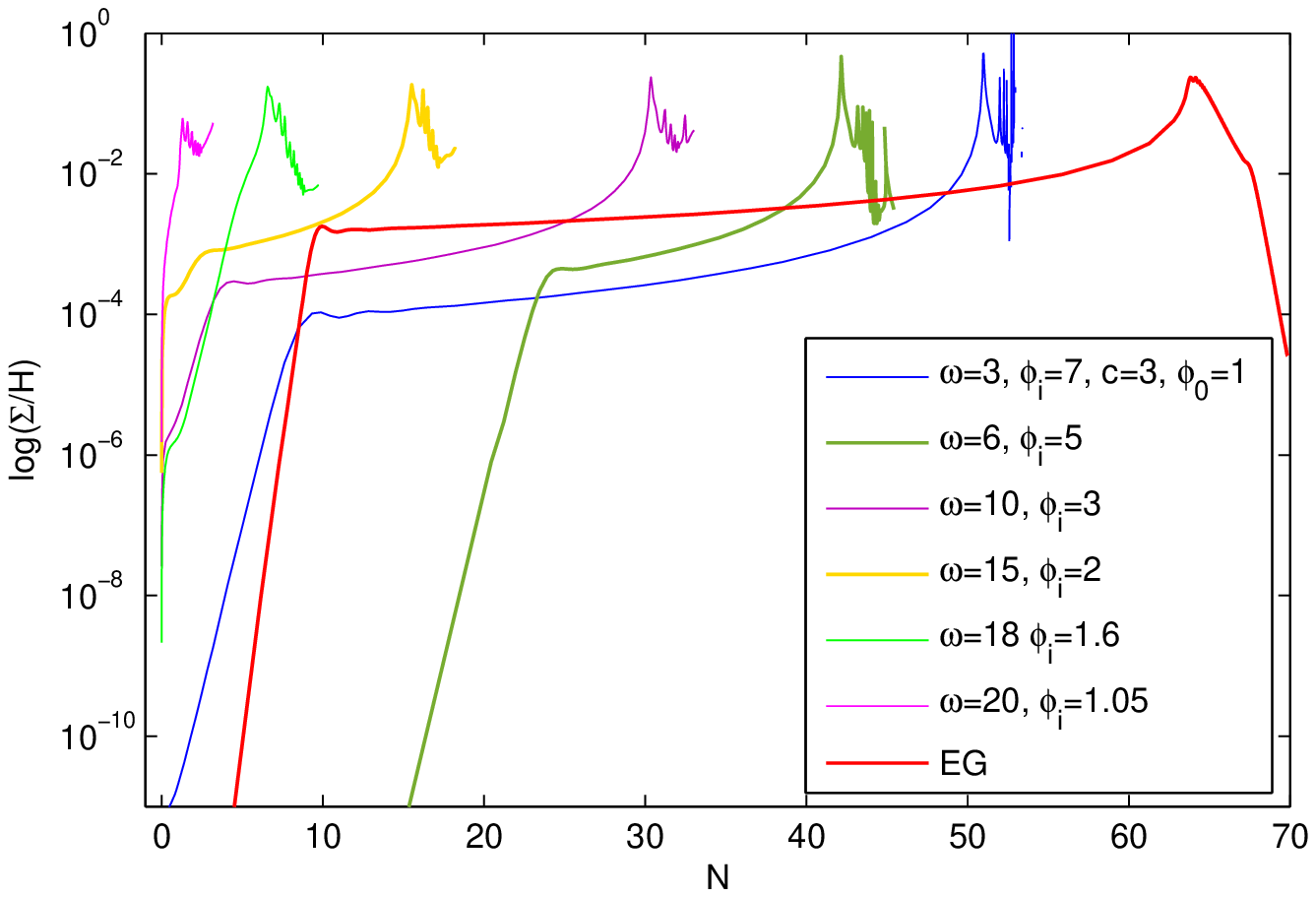} \centering
   \caption{ The same as Fig. \ref{2anisotropyOmega0to7r3}, but for $ \phi_0=1 $, constant value of $ c=3 $ and different values of $ \omega $. Anisotropy in the Einstein gravity for the same potential and the same value for $ c $ and $ \phi_0 $ is plotted in red. }
   \label{anisotropyPhi01w3t020r2phi105to7}
    \end{figure}

The case $ c=2 $ and $ \omega=4 $ is shown in Fig. \ref{anisotropyOmega4r2to45phi100to130}. It shows that there is just one anisotropic phase and anisotropy decreases to the order $ \mathcal{O}(10^{-9}) $, which is closer to the bound found by \cite{Komatsu} (they found a bound for anisotropy of order $ \Sigma/H<10^{-9} $). In this case increasing $ \phi_i $ causes anisotropy reduction. Moreover, increasing $ c $, shifts the anisotropy to larger e-folds. Anisotropy in the Einstein gravity for the same potential is plotted in red for comparison. Clearly, the anisotropy in the BD gravity is smaller than the Einstein gravity. 

The case $ \omega=0 $ is considered in Fig. \ref{3anisotropyOmega0r25to13}. For this case the kinetic term in Eq. \eqref{1} vanishes. This is a way to recover $ f(R) $ gravity (which can be regarded as a special case of BD theory with $ \omega_{BD}=0 $)\cite{Chiba}. Hence the potential \eqref{potential} changes into the potential corresponding to Starobinsky $ R^{2} $ inflation in the Einstein frame. Therefore the potential \eqref{potential} in the BD gravity is a generalized version of Starobinsky $ R^{2} $ inflation. In this case as we increase $ c $ anisotropy decreases and starts to oscillate. This oscillation is what we expect from analytical solution i.e. Eq. \eqref{anisotropy3}. From eq. \eqref{anisotropy3} it is clear that as we increase $ c $, anisotropy goes to zero for a constant value of $ \varepsilon_3$­. But $ \varepsilon_3 $­ is not a constant and it depends on $ \phi $. By increasing $ c $, the anisotropy is shifted to bigger values of $ N $. So the scalar field $ \phi $ starts to oscillate near the end of inflation and $ \varepsilon_3 $­ oscillates too. Therefore this oscillation causes oscillation of anisotropy. If we consider upper bound on the number of efolds from CMB i.e. 60 e-folds, anisotropy does not have enough time to exit the horizon and it is suppressed. As we increase $ c $ more than $ 13 $, numerical solution diverges to infinity. So we could choose $ 13 $ as an upper bound for $ c $. 

 In Fig. \ref{anisotropyPhi01w3t020r2phi105to7} we choose $ \phi_0=1 $. Therefore anisotropy starts from a nonzero value. In this case, increasing $ \omega $ for a constant value of $ c $ leads to elevation of the anisotropy and shifts the anisotropy  to smaller e-folds. Thus, the area in which $ \Sigma/H $ is constant decreases, until this area disappears.

\section{Discussion and Conclusions}\label{conclusion}
In the present work, we studied anisotropic inflation in the Brans-Dicke gravity where the quadrupole power asymmetry is modeled by the kinetic part of an abelian gauge field coupled to the inflaton field. We obtained  a relation for the anisotropy $ \Sigma/H $ which is proportional to the slow-roll parameter, namely $ \varepsilon_{3} $. As an example we carried out the numerical calculation for the displaced quadratic potential and we saw that there is a phase transition in this model, anisotropy grows with $ N $ and depending on the $ \omega $ and $ c $, it can exit the horizon. As we increase $ \omega $, depending on $ \phi_{0} $ in the inflation potential, attractor solution may occur in larger or in smaller values $ \phi_i $. For the case that, attractor solution occurs in smaller  $ \phi_i $, anisotropy is shifted to smaller efolds until it is disappeared. This is because $ \phi_{i} $ converges to $ \phi_{0} $, and consequently $ \varepsilon_3 $ goes to zero. Thus,  $ \Sigma/H $ is disappeared  for larger values of $ \omega $. In the other case, by increasing $ \omega $, we should increase $ \phi_{i} $ to have attractor solution. Increasing $ \phi_{i} $ shifts the anisotropy to larger e-folds until it leaves $ 60 $ e-folds. We also considered the case $ \omega=0 $, where the displaced quadratic inflationary potential is changed to the potential corresponding to Starobinsky $ R^{2} $ inflation in the Einstein frame. Moreover, increasing $ c $ causes the anisotropy to shift to bigger values of $ N $. So anisotropy decreases and starts to oscillate. As a special case, $ c=2 $ was considered. In this case, anisotropy is decreased significantly depending on $ \omega $. For $ \omega=4 $, anisotropy is decreased to the order $ \mathcal{O}(10^{-9}) $.

\acknowledgments
M. T. would like to thank Abolhassan Mohammadi for his help and Kazem Rezazadeh for useful discussion.










\end{document}